\documentclass[aps,pra,twocolumn,showpacs,preprintnumbers,amsmath,amssymb,superscriptaddress,10pt]{revtex4-1}

\usepackage{amsmath,amssymb}
\usepackage{bm}
\usepackage[latin1]{inputenc}
\usepackage{dcolumn}
\usepackage{graphicx}

\newcommand{\Tr}[0]{\text{Tr}}

\begin{document}

\title{Secure gated detection scheme for quantum cryptography}

\author{Lars Lydersen}
\email{lars.lydersen@iet.ntnu.no}
\affiliation{Department of Electronics and Telecommunications, Norwegian University of Science and Technology, NO-7491 Trondheim, Norway}
\affiliation{University Graduate Center, NO-2027 Kjeller, Norway}

\author{Vadim Makarov}
\affiliation{Department of Electronics and Telecommunications, Norwegian University of Science and Technology, NO-7491 Trondheim, Norway}

\author{Johannes Skaar}
\affiliation{Department of Electronics and Telecommunications, Norwegian University of Science and Technology, NO-7491 Trondheim, Norway}
\affiliation{University Graduate Center, NO-2027 Kjeller, Norway}

\date{\today}

\begin{abstract}
Several attacks have been proposed on quantum key distribution systems with gated single-photon detectors. The attacks involve triggering the detectors outside the center of the detector gate, and/or using bright illumination to exploit classical photodiode mode of the detectors. Hence a secure detection scheme requires two features: The detection events must take place in the middle of the gate, and the detector must be single-photon sensitive. Here we present a technique called \emph{bit-mapped gating,} which is an elegant way to force the detections in the middle of the detector gate by coupling detection time and quantum bit error rate. We also discuss how to guarantee single-photon sensitivity by directly measuring detector parameters. Bit-mapped gating also provides a simple way to measure the detector blinding parameter in security proofs for quantum key distribution systems with detector efficiency mismatch, which up until now has remained a theoretical, unmeasurable quantity. Thus if single-photon sensitivity can be guaranteed within the gates, a detection scheme with bit-mapped gating satisfies the assumptions of the current security proofs.
\end{abstract}

\maketitle

\section{Introduction}
Quantum mechanics allows two parties, Alice and Bob, to grow a random, secret bit string at a distance \cite{bennett1984,ekert1991,bennett1992,gisin2002}. In theory, the quantum key distribution (QKD) is secure, even if an eavesdropper Eve can do anything allowed by the currently known laws of nature \cite{mayers1996,mayers2001,lo1999,shor2000,scarani2009}.

In practical QKD systems there will always be imperfections. The security of QKD systems with a large variety of imperfections has been proved \cite{mayers1996,inamori2007,koashi2003,fung2009,lydersen2010,gottesman2004,maroy2010}. Device-independent QKD tries to minimize the number of assumptions on the system, but unfortunately the few assumptions \cite{ekert1991,barrett2005,acin2007} in the security proofs seem to be too strict to allow useful implementations \footnote{There exists at least one proposal for implementing device-independent QKD \cite{gisin2010}. The implementation looks challenging and leads to a much lower secret key rate than the rate in conventional systems.} with current technology \cite{scarani2009a}.

Several security loopholes caused by imperfections have been identified, and attacks have been proposed and in some cases implemented \cite{vakhitov2001,gisin2006,makarov2006,qi2007,lamas-linares2007,fung2007,sauge,makarov2009,nauerth2009,lydersen2010,lydersen2010a,zhao2008,xu2010,gerhardt2010}. With notable exceptions \cite{vakhitov2001,gisin2006,nauerth2009,fung2007,xu2010}, most of the loopholes are caused by an insufficient model of the detectors.

While several detection schemes exist, most implementations use avalanche photodiodes (APDs) gated in the time-domain to avoid high rate of dark counts. Gated means that the APD is single-photon sensitive only when a photon is expected to arrive, in a time window called the \emph{detector gate}. Attacks on these detection schemes are based on exploiting the classical photodiode mode of the APD, or the detector response at the beginning/end of the detector gate.

In the attacks based on the classical photodiode mode of the APD, the detectors are triggered by bright pulses \cite{sauge,lydersen2010a}. If necessary, the APDs can be kept in the classical photodiode mode, in a so-called \emph{blind} state, using additional bright background illumination \cite{sauge,makarov2009,lydersen2010a,lydersen2010b,gerhardt2010}. When the detectors are blind, they are not single-photon sensitive any more, but only respond to bright optical trigger pulses. In most gated systems, blinding is not necessary because the APDs are in the classical photodiode mode outside the gates. Therefore, in the \emph{after-gate attack} \cite{wiechers2010}, the trigger pulses are simply placed after the gate.

\begin{figure}[t!]
  \includegraphics[width=6.95cm]{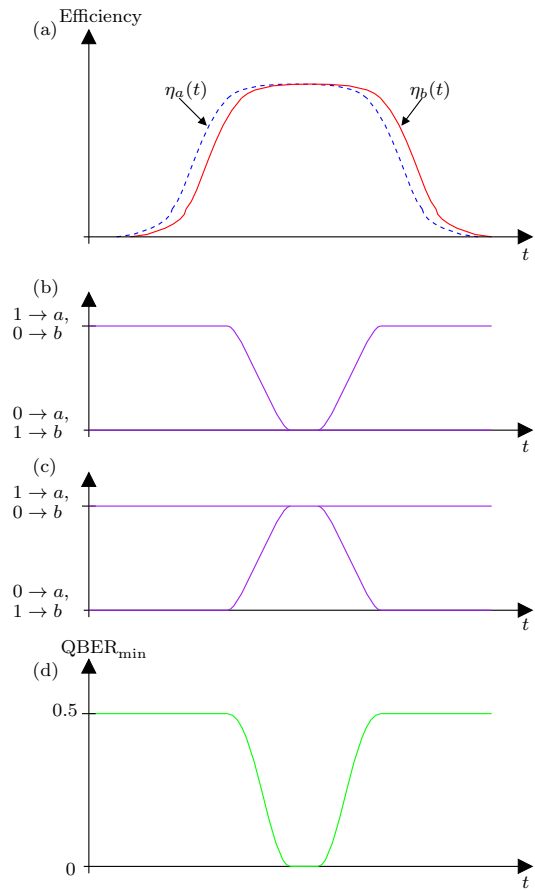}
  \caption{(Color online) Bit-mapped gating. (a) Detector gates with DEM. $\eta_a(t)$ (blue, dashed) and $\eta_b(t)$ (red, solid) are the efficiencies of the two detectors $a$ and $b$. (b),(c) Possible optical bit-mapping (purple) when the software bit-mapping is set to $a \to 0$, $b \to 1$ (Fig.~(b)) and $a \to 1$, $b \to 0$ (Fig.~(c)). In a phase-encoded system the two levels would correspond to $0$ and $\pi$ phase shift in one basis, and $\pi/2$ and $3\pi/2$ phase shift in the opposite basis. Note that the software bit-mapping and the optical bit-mapping coincide in the bit-mapped gate, which is well within the detector gates. (d) $\text{QBER}_{\min} (t)$ (green) as obtained from \eqref{eq:minimum_QBER} with the bit-mapped gate shown in (b) and (c).}
  \label{fig:time-diagram-2}
\end{figure}

Several attacks are based on \emph{detector efficiency mismatch} (DEM) \cite{makarov2006}. If Bob's apparatus has DEM, Eve can control the efficiencies of Bob's detectors individually, by choosing a parameter $t$ in some external domain. Examples of such domains can be the timing, polarization, or frequency of the photons \cite{makarov2006,fung2009}. As an example, consider DEM in the time-domain. Usually Bob's apparatus contains two single-photon detectors to detect the incoming photons, one for each bit value. Due to different optical path lengths, inaccuracies in the electronics, and finite precision in detector manufacturing, the detection windows and hence the efficiency curves of the two detectors $a$ and $b$ are slightly shifted, as seen in Fig.~\ref{fig:time-diagram-2}(a). Several attacks exploit DEM \cite{makarov2006,qi2007,lydersen2010} in various protocols \cite{makarov2008}, some of which are implementable with current technology. The time-shift attack \cite{qi2007} has been used to gain an information-theoretical advantage for Eve when applied to a commercially available QKD system \cite{zhao2008}. In the experiment, Eve captured partial information about the key in 4\% of her attempts, such that she could improve her search over possible keys.

After each loophole has been identified, effort has been made to restore security of the detection schemes. DEM is now included in the receiver model of several security proofs \cite{fung2009,lydersen2010,maroy2010} as an efficiency mismatch or blinding parameter $\eta$, defined differently according to the generality of the proof. For arbitrary systems that can be described with linear optics \cite{lydersen2010},
\begin{equation}
   \eta = \frac{\min_t \left\{ \eta_a(t), \eta_b(t)\right\}}{\max_t \left\{ \eta_a(t), \eta_b(t) \right\}},
   \label{eq:gen_eta}
\end{equation}
where $\eta_a(t)$ and $\eta_b(t)$ are the detection efficiencies of the two detectors. Here $t$ labels the different optical modes; in the special case without mode coupling it labels the different temporal modes. An example is given in Fig.~\ref{fig:time-diagram-2}(a). In the most general case $\eta$ is given by the lowest probability that a non-vacuum state incident to Bob is detected \cite{maroy2010}. For either definition of $\eta$, there is an infinite number of modes involved (all superpositions of temporal modes \cite{lydersen2010}) which makes the blinding parameter difficult to measure or bound in practice. For a given value of $\eta$, the secret key rate is given by \cite{maroy2010}
\begin{equation}
   R \geq -h(E) + \eta (1-h(E)),
   \label{eq:rate_maroy}
\end{equation}
where $E$ is the quantum bit error rate (QBER) measured by Alice and Bob, and $h(\cdot)$ is the binary Shannon entropy function. Here we have assumed symmetry between the bases in the protocol; in addition, we have ignored any basis leakage from Alice and back-reflection from Bob (the most general expression is given in the original reference \cite{maroy2010}). Unfortunately, in practical systems the rate \eqref{eq:rate_maroy} will usually be zero, since $\eta \to 0$ due to the edges of the detector gates. For the commercial QKD system subject to the time-shift attack, $\eta < 0.01$ (estimated from the curves in \cite[Fig.~3]{zhao2008} using Eq.~\eqref{eq:gen_eta}).

\vspace{1 mm}

As noted in \cite{lydersen2010}, one way of obtaining a better $\eta$ would be to discard pulses near the edge of the detector gate. Then $\eta$ could be calculated from \eqref{eq:gen_eta} including only the modes $t$ which are accepted as valid detections. However, this is highly non-trivial. The avalanche in an APD is a random process, and the jitter in the photon-timing resolution is of the same order of magnitude as the duration of the detector gate. A good photon-timing resolving detector still has 27~ps jitter \cite{cova2004}. Furthermore, the unavoidable difference in the acceptance windows for the different detectors will also contribute to DEM (one detector accepts clicks while the other discards them).

\vspace{1 mm}

A frequently mentioned countermeasure for systems with DEM is called \emph{four-state Bob} \cite{nielsen2001,lagasse2005,makarov2006,qi2007}. Then Bob uses a random detector--bit mapping, randomly assigning the bit values 0 and 1 to the detectors $a$, $b$ for each gate. In a phase-encoded QKD system, this can be implemented by Bob choosing from four different phase settings $\left\{0,\pi/2,\pi,3\pi/2\right\}$ instead of only two $\left\{0,\pi/2\right\}$. Then Eve does not know which detector characteristics correspond to which bit value. However, as mentioned previously \cite{makarov2006,qi2007,lydersen2010} this patch opens a different security loophole. Eve may use a \emph{Trojan-horse attack} \cite{ribordy2000,bethune2000,vakhitov2001,gisin2006} to read Bob's phase modulator settings, thus additional hardware modifications are required. Note also that the four-state Bob patch does not secure against the after-gate attack \cite{wiechers2010} nor any of the detector control attacks \cite{lydersen2010a,lydersen2010b}.

Here we present a novel way of securing Bob's receiver called \emph{bit-mapped gating} (Section~\ref{sec:basis_gating}). It secures the system against all kinds of pulses outside the central part of the detector gate in the Bennett-Brassard 1984 (BB84) and related protocols \cite{bennett1984,hwang2003,lo2005,wang2005a}. The technique is compatible with the existing security proofs \cite{fung2009,lydersen2010,maroy2010} and makes it simple to find $\eta$. In general it represents a useful concept, where parameters from characteristics of the QKD system are coupled to the parameters estimated by the protocol. In this case $\eta$ becomes coupled to the QBER. Subsequently we analyze the security of bit-mapped gating (Section~\ref{sec:security_analysis}), discuss how to characterize detectors, and how to implement a guarantee of single-photon sensitivity (Section~\ref{sec:detector-design}). Finally we conclude (Section~\ref{sec:discussion}).

\section{Bit-mapped gating}
\label{sec:basis_gating}
Let us start with two definitions. The \emph{software bit-mapping} determines how the signals from detectors $a$ and $b$ are mapped into the logical bits $0$, $1$. Similarly the \emph{optical bit-mapping} which can be implemented by generalizing the basis selector, maps quantum states with bit values $0$, $1$ (for instance $|0\rangle$, $|1\rangle$ in the $Z$-basis) to the detectors $a$, $b$. Note that if the software bit-mapping and the optical bit-mapping do not coincide, a bit value $0$ sent by Alice will be detected as bit value $1$ by Bob.

\emph{Bit-mapped gating} works as follows: 
\begin{itemize}
  \item Somewhere in between the detector gates, Bob randomly selects the software bit-mapping, assigning detectors $a$, $b$ to bit values $0$, $1$. 
  \item Likewise, the basis is selected randomly between the $X$ and $Z$ basis, along with a random optical bit-mapping. Since this happens between the detector gates, jitter is not critical.
  \item Inside the detector gate, the optical bit-mapping is matched to the software bit-mapping. The period with matching optical and software bit-mapping is the \emph{bit-mapped gate.}
\end{itemize}
Note that the optical bit-mapping can be equal on both sides of the bit-mapped gate to minimize the need for random numbers. Fig.~\ref{fig:time-diagram-2} shows a typical time diagram.

As an example, consider a phase-encoded implementation of the BB84 protocol, where the basis selector at Bob is usually a phase modulator. 0 phase shift corresponds to $Z$ basis and $\pi/2$ phase shift corresponds to the $X$ basis. The optical bit-mapping can be selected by adding either $0$ or $\pi$ to the phase shift. Hence in this implementation the bit-mapped gating patch could be implemented as follows: Bob randomly selects the software bit-mapping somewhere between the gates. Furthermore, Bob selects a random basis, i.e$.$ $0$ or $\pi/2$ phase shift between the gates, and adds either $0$ or $\pi$ to the phase shift to apply the random optical bit-mapping. During the gate, the software and the optical bit-mapping coincide.

All states received and detected outside the bit-mapping gate cause random detection results (due to the random optical and software bit-mapping), and thus introduce a QBER of 50\%. The measured QBER could be used to estimate the fraction of detections which must have happened in the center of the gate (in Fig.~\ref{fig:time-diagram-2}: close to zero QBER would mean that most detection events must have passed the basis selector, and thus hit the detector, in the middle of the gate). This can be used to limit the DEM, because considering only the modes in the center of the detector gate gives less DEM than considering all modes.

\section{Security analysis}
\label{sec:security_analysis}
The goal of this section is to derive an expression for the minimum QBER introduced by any state received by Bob, during the transition to and from the bit-mapped gate. Ideally, the minimum QBER is 0 inside the bit-mapped gate, and $1/2$ outside the bit-mapped gate. 

The input of Bob's detection system consists of many optical modes $t$, for instance corresponding to different arrival times at Bob's system. Each mode $t$ may contain a mixture of different number states. Note that Bob could have measured the photon number in each mode without disturbing the later measurement; thus it suffices to address specific number states. We use the usual assumption that each photon in a $n$-photon state is detected individually. Under these assumptions, we first calculate the minimum QBER caused by a single photon arriving in a single mode at Bob. Then, in appendix~\ref{sec:multiphotons} we show that multiple photons in this mode, or photons in other modes can only increase the minimum QBER.

Consider a single photon arriving at Bob in a given mode $t$. Since the BB84 protocol is symmetric with respect to the bit values and the bases, we may assume without loss of generality that Alice sent $Z0$ and that Bob measures in the $Z$ basis. Outside the bit-mapped gate, Bob performs four different measurements depending on the software and optical bit-mapping. For each measurement, Bob will obtain one out of three measurement outcomes, bit 0, bit 1 or vacuum denoted by subscript $v$. 

Let $\eta_a,\eta_b$ be the efficiencies of the two detectors, $|\theta\rangle = \cos\theta |0\rangle + \sin\theta |1\rangle$ and $|\theta^\bot\rangle = \sin\theta |0\rangle - \cos\theta |1\rangle$. During a bit-mapped gate, $\theta$ is varied from $0$ to $\pi/2$. For each value of $\theta$, Bob performs one out of the four measurements
\begin{subequations}
  \begin{equation}
    \begin{split}
      &M_0 = \eta_a |0\rangle \langle0| , M_1 = \eta_b |1\rangle \langle1|, \\
      &M_v = I - M_0 - M_1 \text{,}
    \end{split}
  \end{equation}
  \begin{equation}
    \begin{split}
      &M'_0  = \eta_b |0\rangle \langle0| , M'_1 = \eta_a |1\rangle \langle1|, \\
      &M'_v = I - M'_0 - M'_1 \text{,}
    \end{split}
  \end{equation}
  \begin{equation}
    \begin{split}
      &M''_0  = \eta_a |\theta\rangle \langle\theta| , M''_1 = \eta_b |\theta^\bot\rangle \langle\theta^\bot|,\\
      &M''_v = I - M''_0 - M''_1 \text{,}
    \end{split}
  \end{equation}
  \begin{equation}
    \begin{split}
      &M'''_0  = \eta_b |\theta\rangle \langle\theta| , M'''_1 = \eta_a |\theta^\bot\rangle \langle\theta^\bot|, \\
      &M'''_v = I - M'''_0 - M'''_1 \text{.}
    \end{split}
  \end{equation}
  \label{eq:single_M}
\end{subequations}

If Bob uses the four measurements with equal probabilities, the statistics will be given by using the measurement operators
\begin{subequations}
  \begin{equation}
    \begin{split}
       E_0 &= \frac{1}{4} \left( M_0 + M'_0 + M''_0 + M'''_0 \right) \\
           &= \frac{1}{4}(\eta_a + \eta_b)( (1+\cos^2\theta)|0\rangle\langle0| + \sin^2\theta|1\rangle \langle1|\\
          &+ \sin\theta\cos\theta (|0\rangle\langle1| + |1\rangle\langle0|) )\text{,}
    \end{split}
  \end{equation}
  \begin{equation}
    \begin{split}
      E_1 &= \frac{1}{4} \left( M_1 + M'_1 + M''_1 + M'''_1 \right) \\
          &= \frac{1}{4}(\eta_a + \eta_b)( \sin^2\theta |0\rangle\langle0| + (1+\cos^2\theta )|1\rangle \langle1|\\
          &- \sin\theta\cos\theta (|0\rangle\langle1| + |1\rangle\langle0|) )\text{,}
    \end{split}
  \end{equation}
  \begin{equation}
    \begin{split}
      E_v &= \frac{1}{4} \left( M_v + M'_v + M''_v + M'''_v \right) \\
          &= \left( 1 - \frac{\eta_a + \eta_b}{2} \right) I \text{.}
    \end{split}
  \end{equation}
  \label{eq:single_E}
\end{subequations}
Note that $E_v \propto I$, so the detection probability is independent of the photon-state $\rho$:
\begin{equation}
   p_{\text{det}} = 1 - \Tr[\rho E_v] = \frac{\eta_a + \eta_b}{2} \text{.}
\end{equation}

The eigenvalues of operators $E_0$ and $E_1$ are given by $p_{\text{det}}(1 \pm \cos\theta)/2$. Thus the minimum and maximum probability of detecting bit values 0 and 1 for any single photon sent by Eve is given by
\begin{align}
     p_{0,\min} &= p_{1,\min} = \frac{p_{\text{det}}}{2} (1 - \cos\theta) \text{,} \\
     p_{0,\max} &= p_{1,\max} = \frac{p_{\text{det}}}{2} (1 + \cos\theta) \text{.}
\end{align}
Since Alice sent $Z0$, the minimum QBER introduced by a single photon is given by
\begin{equation}
   \text{QBER}_{\min} = \frac{p_{1,\min}}{p_{\text{det}}} = \frac{1}{2} (1 - \cos\theta)
   \label{eq:minimum_QBER}
\end{equation}
As expected, for $\theta = \pi/2$ $\text{QBER}_{\min} = 1/2$. For multiphotons, a random bit value is assigned to double clicks \cite{gottesman2004,inamori2007}.  Appendix~\ref{sec:multiphotons} shows that sending multiple photons can only increase the QBER caused by detection events. Hence Eq.~\eqref{eq:minimum_QBER} gives the minimum QBER for any photonic state sent by Eve.

The security proofs in Refs.~\cite{fung2009,lydersen2010,maroy2010} involve Bob predicting the results of Alice's virtual $X$-basis measurement. Since the prediction is not carried out in practice, Bob can perform any operation permitted by quantum mechanics. In the proofs Bob's prediction consists of a filter followed by an ``X-basis'' measurement. When nothing is known about the distribution of the detection events within the gate, the worst case assumption is that all the detection events occur with maximum DEM. Therefore, the best filter we can construct can only guarantee that a fraction $\eta$ of the inputs can successfully pass the filter.

\begin{figure}[t!]
  \includegraphics[width=6.5cm]{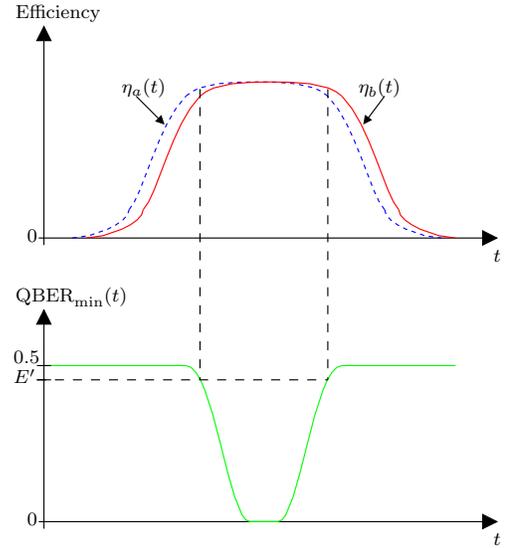}
  \caption{(Color online) Curves (a) and (d) from Fig.~\ref{fig:time-diagram-2}. The dashed line shows how a threshold $E'$ can be used to limit the range of modes $t$ used to calculate or bound $\eta'$.}
  \label{fig:time-diagram-3}
\end{figure}

With our patch, we may use the QBER to determine a lower bound for the number of detection events which must have happened in the central part of the detector gate. Assuming that $t$ labels temporal modes, consider the number of detection events which occurred in the range where $\text{QBER}_{\min}< E'$ (see Fig.~\ref{fig:time-diagram-3}). Here, $E'$ is a threshold selected by Bob. Let $\eta'$ be the blinding parameter for the modes for the range where $\text{QBER}_{\min}< E'$. It can be calculated from Eq.~\eqref{eq:gen_eta}, but where $t$ only runs over this range. If the measured QBER is equal to $E$, a fraction 
\begin{equation}
    f=\frac{E' - E}{E'},
\end{equation}
must have been detected in the modes where $\text{QBER}_{\min}< E'$. Note that increasing $E'$ increases $f$, and may decrease $\eta'$ (see Fig.~\ref{fig:time-diagram-3}). As will become apparent below, $E'$ should be selected to maximize $f\eta'$.

For decoy protocols \cite{hwang2003,lo2005,wang2005a}, $E$ should be replaced with the QBER estimated for single-photon states. This improves the estimate of the fraction $f$, especially for large distances where the dark counts become a major part of the total QBER.

In the worst case, a fraction $f$ experienced a reduced DEM $\eta'$. Therefore, the filters in the security proofs can be replaced as follows: the new filter discards pulses in the modes for which $\text{QBER}_{\min}> E'$. For the modes inside the bit-mapped gate, where $\text{QBER}_{\min}< E'$, the new filter reverts the quantum operation from the receiver in the opposite basis in the same way that the old filter reverted it for all modes, but now having success rate $\eta'$. Since we can guarantee that a fraction $f$ of the photons are in the bit-mapped gate, at least $\eta'f$ pulses will successfully pass the new filter. Therefore the parameter $\eta$ in all the proofs \cite{fung2009,lydersen2010,maroy2010} can be replaced with $\eta'f$, and the rate \eqref{eq:rate_maroy} becomes
\begin{equation}
   R \geq -h(E) + f\eta'[1-h(E)],
   \label{eq:new_rate}
\end{equation}
when one assumes symmetry between the bases, and no source errors. Without symmetry between the bases, all parameters become basis-dependent, and the rate is the sum of the rates in each basis.

Let us see how bit-mapped gating could improve the secure key rate for the commercial QKD system in \cite{zhao2008}. For this system $\eta < 0.01$. In the same experiment, the QBER is measured to be 5.68\%. Assuming $E' = 0.45$ and $\eta' = 0.9$, $f\eta'$ becomes $0.79$; thus a substantial improvement. In fact, the rate obtained from Eq.~\eqref{eq:rate_maroy} without the patch is 0, while the rate obtained from Eq.~\eqref{eq:new_rate} is 0.227, so clearly the patch can be used to re-secure an insecure implementation.

\section{Detector design and characterization}
\label{sec:detector-design}
When designing Bob's system, one should ensure that the bit-mapped gate is well within the detector gate, i.e. that the detector efficiencies are approximately equal within the bit-mapped gate. Then, it should be possible to measure or bound the detector efficiencies and the basis selector response $\theta(t)$ in the temporal domain. In a phase-encoded system this would correspond to measuring the detector efficiencies and the phase modulation as a function of time \footnote{If the phase modulator response differs depending on the software bit-mapping and basis choice, it should simply be bounded.}, over the range of wavelengths and polarizations accepted by Bob. With this data, the minimum QBER as a function of time can be calculated from \eqref{eq:minimum_QBER}, and a diagram similar to Fig.~\ref{fig:time-diagram-3} can be obtained. After selecting an appropriate limit $E'$, $\eta'$ can be calculated by \eqref{eq:gen_eta} but where $t$ runs only over the modes where $\text{QBER}_{\min}< E'$, and not over all available modes.

In general there might be coupling between the different temporal modes due to misalignments and multiple reflections \cite{fung2009,lydersen2010}. The bit-mapped gate ensures that the pulse passed the basis selector inside the temporal detector gate, but does not guarantee the actual detection time. For example, a pulse could pass in the center of the bit-mapped gate, but afterwards take a multiple reflection path such that it hits the detector outside the detector gate. This can be handled by characterizing the worst case mode coupling as described previously \cite{lydersen2010}. Let $\delta$ be the worst case (power) coupling of modes inside the bit-mapped gate to outside the gate. This will typically be the worst case multiple-reflection path after the basis selector, and should be boundable from component characteristics. Then, the parameter $\delta$ can be interpreted as
\begin{equation}
   \delta = \frac{\text{\#pulses that hits the detector outside the gate}}{\text{\#pulses sent into the gate}}.
\end{equation}
In the worst case, $\delta$ of the $f$ detection events might have happened outside the central part of the detector gate; thus one must let $f \to f(1-\delta)$.

Finally one must guarantee that the detectors are not blind within the gate \cite{lydersen2010a}, and fulfill the assumptions in Section~\ref{sec:security_analysis} during the transition of the optical bit-mapping. Note that the transition ends when there is no longer any correlation between the software bit-mapping and the optical bit-mapping. If a significant correlation exists also after the detector gate, it could be exploited in the after-gate attack \cite{wiechers2010}.

Although it is tempting to place an optical watchdog detector at the entrance of Bob, the absence of bright illumination does not necessarily mean that the detectors are single-photon sensitive. For instance, due to the thermal inertia of the APD, it can remain blind for a long time after the bright illumination is turned off \cite{lydersen2010b}.

A cheap way to guarantee single-photon sensitivity is to monitor all detector parameters \cite{makarov2009}, such as APD bias voltage, current and temperature. It seems difficult to monitor the temperature of the APD chip \cite{lydersen2010b}, but monitoring the bias voltage and current should make it possible to predict the heat generated by the APD, and thus prevent thermal blinding \cite{lydersen2010b}.

\begin{figure}[t!]
  \includegraphics[width=6.5cm]{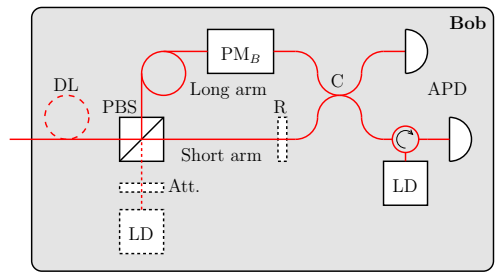}
  \caption{(Color online) A calibrated light source inside Bob. The figure shows the Bob module in a plug-and-play system \cite{muller1997,zbinden1997,ribordy1998,gisin2002} with two possible implementations of the calibrated light source: either a separate attenuated laser diode (LD) at a suitable place, or in the case of send-return systems where Bob already contains a laser diode a weakly reflective element (R) to reflect some light back into the APDs. A short delay line (DL, $\text{delay} > \text{gate period} / 2$) at Bob's input guarantees that Eve cannot interfere with the detector operation based on whether the source is activated or not. PBS: polarizing beam splitter; Att.: optical attenuator; PM: phase modulator; C: 50/50\% fiber-optic coupler. }
  \label{fig:light-source-inside-bob}
\end{figure}

The ultimate way of guaranteeing single-photon sensitivity is to measure it directly. This can be done by placing a calibrated light source inside Bob that emits faint pulses at random times \cite{gerhardt2010} (see Fig.~\ref{fig:light-source-inside-bob}). Then the absence of detection events caused by this source
would indicate that the detector is blind. Further, a calibrated light source inside Bob could be useful in more ways, for instance to characterize and calibrate detector performance in deployed systems.

The patch could cause a minor reduction in QKD performance compared to running an (insecure) system without the patch. In particular, the detector gates might have to be longer to contain the basis-selector gate. This would increase the dark count rate, and thus limit the maximum transmission distance. A calibrated light source inside Bob would also cause a minor reduction in the performance since the gates used for testing the detector sensitivity likely cannot be used to extract the secret key. However, both these effects are minor, and are easily justified by the restoration of security.

\section{Discussion and conclusion}
\label{sec:discussion}
In this work, we have presented a technique called ``bit-mapped gating'' to secure gated single-photon detectors in QKD systems. It is based on a general concept where hardware imperfections are coupled to the parameters estimated by the protocol. Bit-mapped gating causes all detection events outside the central part of the detector gate to cause high QBER.

Bit-mapped gating is compatible with the current security proofs for QKD systems with detector efficiency mismatch \cite{fung2009,lydersen2010,maroy2010}. In particular it provides a simple way of measuring the detector blinding parameter. A secure gated detection scheme is obtained if bit-mapped gating is combined with detectors guaranteed to be single-photon sensitive.

\begin{acknowledgments}
Financial support is acknowledged from the Research Council of Norway (grant no$.$ 180439/V30)
\end{acknowledgments}

\appendix
\section{Minimum QBER for multiphotons}
\label{sec:multiphotons}
Here we prove that the minimum QBER can only increase when the number of photons sent to Bob is increased. As noted previously we use the usual assumption that each photon in a $n$-photon state is detected individually. This means that each photon hits a separate set of detectors, and then the detection results are merged to give the detection results of threshold detectors.

Let us first consider the case where Bob receives a large number of two-photon states. Let the two photons within the states be labeled 1 and 2. Individually, each of the two photons would have caused the minimum QBER $Q_1$ and $Q_2$ (as found from Eq.~\eqref{eq:minimum_QBER}). Again we assume that Alice sends the bit value 0, without loss of generality. For two-photon states there will be three cases of detected events: either only photon 1 is detected, only photon 2 is detected, or both photons are detected (in our model, this latter possibility corresponds to the case where both sets of detectors register a click). Let there be $n_1$ events where only photon 1 was detected, $n_2$ events where only photon 2 was detected, and $c$ events where both photons were detected. For photon $i$, out of the $n_i = n_{i,0} + n_{i,1}$ events, $n_{i,0}$ and $n_{i,1}$ were detected as the bit value 0 and 1, respectively. Likewise, out of the $c = c_{i,0} + c_{i,1}$ events where both photons are detected, $c_{i,0}$ and $c_{i,1}$ were detected as the bit value 0 and 1 for photon $i$ (remember that in the model each photon hits a separate set of detectors).

When only one of the photons is detected, the situation is identical to the single-photon case treated in Section~\ref{sec:security_analysis}. Hence states such that $Q_i = n_{i,1}/n_i$ give the lowest possible QBER. For the events where both photons are detected, the detections can have any correlation, but for each photon $c_{i,1} \geq cQ_i$ since $Q_i$ represents the lowest fraction of the bit value 1 possible, regardless of the correlation with any other photon. The total QBER $Q$ can be found from merging the detections from the two sets of detectors. Double clicks are assigned a random bit value \cite{gottesman2004,inamori2007}, therefore half of the double clicks get the bit value 1. This gives the total QBER
\begin{equation}
  \begin{split}
   Q &= \frac{n_{1,1} + n_{2,1} + \frac{1}{2}\left(c_{1,1} + c_{2,1}\right)}{n_1 + n_2 + c} \\
     &\geq \frac{Q_1 \left(n_1 + \frac{c}{2} \right) + Q_2 \left(n_2 + \frac{c}{2}\right)}{n_1 + n_2 + c} \\
     &\geq \min \left( Q_1, Q_2 \right).
  \end{split}
\end{equation}

By repeating the argument above, but replacing the detection of photon 1 with the detection of $N$ photons, it is easy to see that $Q \geq \min \left( Q_{N}, Q_{N+1} \right)$. Hence by induction, any detection event caused by more than one photon can only cause a higher QBER than the single-photon case.

\end{document}